# Solar energy harvesting with carbon nitrides:

# Do we understand the mechanism?

Wolfgang Domcke[1], Johannes Ehrmaier[1] and Andrzej L. Sobolewski[2]

[1] Department of Chemistry, Technical University of Munich, D-85747 Garching, Germany

[2] Institute of Physics, Polish Academy of Sciences, 02-668 Warsaw, Poland

Email: domcke@ch.tum.de  (Wolfgang Domcke)

**Abstract**

The photocatalytic splitting of water into molecular hydrogen and molecular oxygen with sunlight is the dream reaction for solar energy conversion. Since decades, transition-metal-oxide semiconductors and supramolecular organometallic structures have been extensively explored as photocatalysts for solar water splitting. More recently, polymeric carbon nitride materials consisting of triazine or heptazine building blocks have attracted considerable attention as hydrogen-evolution photocatalysts. The mechanism of hydrogen evolution with polymeric carbon nitrides is discussed throughout the current literature in terms of the familiar concepts developed for photoelectrochemical water splitting with semiconductors since the 1970s. We discuss in this perspective an alternative mechanistic paradigm for photoinduced water splitting with carbon nitrides, which focusses on the specific features of the photochemistry of aromatic N-heterocycles in aqueous environments. It is shown that a water molecule which is hydrogen-bonded to an N-heterocycle can be decomposed into hydrogen and hydroxyl radicals by two simple sequential photochemical reactions. This concept is illustrated by first-principles calculations of excited-state reaction paths and their energy profiles for hydrogen-bonded complexes of pyridine, triazine and heptazine with a water molecule. It is shown that the excited-state hydrogen-transfer and hydrogen-detachment reactions are essentially barrierless, in sharp contrast to water oxidation in the electronic ground state, where high barriers prevail. We also discuss in some detail the products of possible reactions of the highly reactive hydroxyl radicals with the chromophores. We hypothesize that the challenge of efficient solar hydrogen generation with carbon-nitride materials is less the decomposition of water as such, but rather the controlled recombination of the photogenerated radicals to the closed-shell products $H_2$ and $H_2O_2$.

# 1. Introduction

The decomposition of water molecules into molecular hydrogen and molecular oxygen with solar photons

$$2\,H_2O\ +\ n\,h\nu\ \rightarrow\ 2\,H_2\ +\ O_2 \qquad\qquad\qquad (1)$$

is the dream reaction for the direct conversion of solar energy into chemical energy. Molecular hydrogen can either directly be used as a fuel in a future global hydrogen economy or can be used for the reduction of carbon dioxide to generate clean liquid fuels, such as methanol or ethanol. Molecular oxygen is the waste product to be released into the atmosphere.

The apparent simplicity of Eq. (1) is deceptive. The fragmentation of $H_2O$ is strongly endothermic and at least two UV/vis photons are required to split an OH bond of water. Since water does not absorb in the spectral range of sunlight at the surface of earth, a material with a suitable band gap is needed which efficiently absorbs sunlight and generates the oxidative and reductive equivalents which are necessary for the decomposition of water.

The extensive world-wide research efforts towards solar water splitting are to a large extent based on two alternative strategies, which may be termed the photoelectrochemical approach and the artificial photosynthesis approach, respectively. Photoelectrochemical water splitting was first demonstrated by Fujishima and Honda in 1972 [1]. Following light absorption, electrons and holes are generated in a semiconductor and are separated by charge migration driven by an external voltage. $H_2$ is assumed to be catalytically formed from electrons and protons at the cathode and $O_2$ is assumed to be catalytically generated by the reaction of holes with water at the anode. Research in recent years has primarily been focused on finding semiconductors with smaller band gaps and on the discovery of more efficient and robust hydrogen-evolution and oxygen-evolution catalysts [2-4]. Despite these extensive efforts over many years, just a few reports of stoichiometric photocatalytic water splitting with visible light and without external bias voltage exist and the quantum efficiency is still low [5-7]. The alternative approach, artificial photosynthesis, is inspired by the structure and function of the natural photosynthetic reaction center photosystem II. In artificial photosynthesis, complex molecular or supra-molecular structures have been devised and synthesized, which typically consist of an organic or organometallic chromophore as photosensitizer, an electron donor as well as an electron acceptor [8-10]. The goal is a fast stepwise unidirectional electronic charge separation over large distances after the excitation of the chromophore, such that charge-recombination processes are minimized. While lifetimes of charge-separated states of the order of microseconds were reported, the efficient coupling of these charge-separation processes with the multi-electron redox processes involved in $H_2$ and $O_2$ formation has not yet been demonstrated [10-12].

In both photoelectrochemical water splitting and in artificial photosynthesis, water molecules are not dissociated directly with photons. The absorbed photons rather generate mobile electrons and holes in semiconductors or in extended molecular donor-acceptor systems. The charges migrate apart and eventually react with water, whereby $H_2$ and $O_2$ are formed in multi-electron redox reactions with suitable catalysts. One of the fundamental challenges in both concepts arises from the substantial charge-recombination losses that occur when charges have to be separated over long distances and long timescales. From a fundamental mechanistic point of view, both paradigms, the photoelectrochemical approach to water splitting and the concept of artificial photosynthesis, appear not really convincing. Charge migration in semiconductors to interfaces at mesoscopic length scales may require nanoseconds to microseconds and the final oxidation of water by holes at interfaces may take even longer, since a high barrier (about 2.0 eV) has to be overcome with the help of water oxidation catalysts. As a consequence, substantial losses of charges through charge-recombination processes are likely to occur, which leads to the low efficiency of water splitting in current experiments.

In this perspective, we outline an alternative paradigm of photoinduced water splitting in which the transport of charges over substantial length scales and time scales is avoided. We aim instead at the direct homolytic decomposition of water molecules into hydrogen and hydroxyl radicals via fast proton-coupled electron transfer (PCET) reactions in excited electronic states of chromophore-water complexes. As has been demonstrated by ab initio electronic-structure calculations, these excited-state PCET reactions can be barrierless or nearly barrierless. Since electrons and protons are the lightest particles in chemistry, they are expected to react on the fastest time scales. The abstraction of a hydrogen atom from a water molecule via a PCET reaction immediately after the absorption of a photon should be possible on femtosecond time scales if the reaction is nearly barrierless. On this time scale, deleterious reactions hardly can compete, which should lead to a drastic reduction of loss processes.

To achieve such fast reactions, the water molecule to be oxidized should be firmly hydrogen bonded to the chromophore before the photon absorption occurs. Denoting the chromophore by A and the hydrogen-bonded complex of A with a water molecule by A-$H_2O$, we envisage a biphotonic reaction scheme according to the equations

$$A + H_2O \rightarrow A\text{-}H_2O \qquad (2)$$

$$A\text{-}H_2O + h\nu \rightarrow A^*\text{-}H_2O \qquad (3)$$

$$A^*\text{-}H_2O \rightarrow AH + OH \qquad (4)$$

$$AH + h\nu \rightarrow A + H. \qquad (5)$$

The overall reaction is

$$H_2O + 2 h\nu \rightarrow H + OH. \qquad (6)$$

In the hydrogen-bonded complex A-$H_2O$, A is the proton acceptor (a base) and $H_2O$ is the proton donor (an acid). Water is oxidized by the reaction (4), while the chromophore A is reduced in this reaction. The reduced chromophore AH is oxidized by the reaction (5). The chromophore A is thereby regenerated and the net reaction is the homolytic splitting of a water molecule into H and OH radicals by the sequential absorption of two photons.

In this perspective, this novel photochemical concept of water splitting is outlined in some detail for organic chromophores, in particular so-called carbon nitrides. In pioneering work in the 1980s and 1990s, Yamagida and coworkers demonstrated light-driven hydrogen evolution with π-conjugated linear polymers, such as polypyridine, in the presence of colloidal noble metals and a sacrificial electron donor [13,14]. More recently, covalent organic frameworks consisting of N-heterocycles, such as *s*-triazine, and aromatic linkers were synthesized and tested for hydrogen evolution with visible light [15-19]. Quasi-two-dimensional polymeric materials consisting of heptazine (tri-*s*-triazine) building blocks connected by nitrogen atoms or imide groups have received enormous attention since the discovery of their photocatalytic activity for hydrogen evolution with UV/vis light [20], see [21-27] for recent reviews. Most of these heptazine-based materials are closely related to Liebig's "melon" [28]. A nano-disperse platinum co-catalyst and a sacrificial electron donor (usually triethanolamine (TEA) or methanol) are required for efficient hydrogen evolution [20-27].

## 2. Hydrogen-bonded chromophore-water clusters

We will consider in this survey the three N-heteroycles pyridine, triazine and heptazine as building blocks of carbon-nitride materials. Owing to the localization of the water-oxidation reaction on a single N-atom of the chromophore, the basic principles of the photochemical reactions can be revealed for hydrogen-bonded complexes of these chromophores with a single water molecule, as displayed in Fig. 1. For brevity and clarity, we consider here only planar structures, that is, the water molecule lies in the plane of the aromatic ring(s). All systems thus exhibit $C_s$ symmetry.

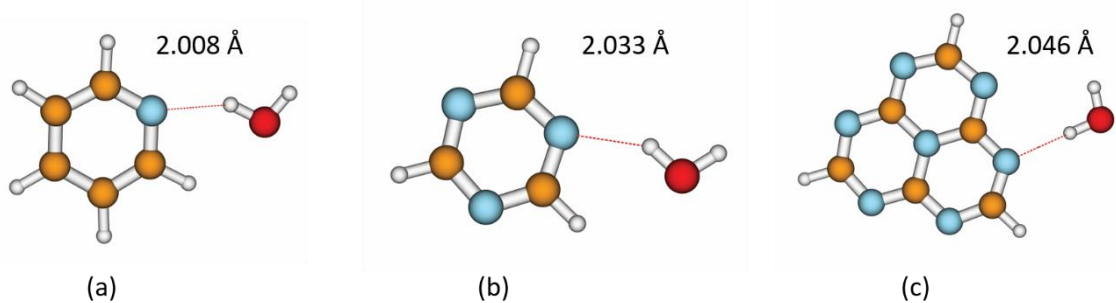

*Fig. 1. Equilibrium geometries of the pyridine-$H_2O$ (a), triazine-$H_2O$ (b) and heptazine-$H_2O$ (c) complexes in the electronic ground state. The length of the hydrogen bond is indicated.*

Pyridine (Fig. 1a) is the smallest aromatic N-heterocycle and represents the simplest model system for the exploration of the photochemistry of carbon nitride materials. It was discovered in the 1990s that the π-conjugated linear polymer polypyridine (PPy) can catalyze hydrogen evolution by UV light in the presence of colloidal noble metals and a sacrificial electron donor [14]. In particular, PPy was found to be a much better photocatalyst than polyphenylene (PP) [13,14] which provided a first hint on the relevance of heterocyclic N-atoms in water photolysis. More recently, the photochemistry of pyridine-$(H_2O)_n$ clusters was investigated by Jouvet and coworkers in a molecular beam [29]. Hydrogen abstraction from water stimulated by 225 nm irradiation was observed via the spectroscopic detection of the pyridinyl (hydrogenated pyridine) radical. This reaction could be detected in clusters with four or more water molecules, albeit not in smaller clusters [29]. Interestingly, related proton-transfer reactions were observed in anionic pyridine-water clusters. It was shown that the most stable form of pyridine-$(H_2O)_n^-$ clusters with $n > 3$ corresponds to a neutral pyridinyl radical which is hydrogen-bonded to an anionic $OH^-(H_2O)_{n-1}$ cluster [30].

Triazine (Fig. 1b) is one of the building blocks of polymeric carbon nitride materials. So-called covalent organic frameworks consisting of triazine units connected by organic linkers were recently synthesized by Lotsch and coworkers and tested for UV/vis-induced hydrogen evolution with sacrificial electron donors. In a systematic study considering pyridine, pyrimidines and *s*-triazine as building blocks of covalent organic frameworks, it was shown that the hydrogen evolution efficiency increases systematically with the nitrogen content of the heterocycles, which underlines the essential role of heterocyclic N-atoms in water-splitting photocatalysis [17,18]. Triazine-based materials consisting of *s*-triazine units connected by N-atoms also have been synthesized and tested as hydrogen-evolution photocatalysts [15,16]. In contrast to the covalent organic frameworks, these materials are pure carbon nitrides. The band structures of these periodic triazine-based materials were investigated with density functional theory (DFT) [31-33]. Butchosa et al. computed the optical properties of triphenyl-triazine oligomers and clusters thereof with time-dependent DFT (TDDFT) and with coupled-cluster (CC2) methods [34]. Srinivasu and Ghosh also studied the adsorption of water molecules and radicals on periodic triazine-based systems and calculated the energy profile for the dissociation of an adsorbed water molecule, which exhibits a high barrier (about 2.0 eV) [35].

Heptazine (Fig. 1c) has been identified by Pauling as the building block of Liebig's "melon" [36]. Although the precise composition and structure of the polymer melon depend on the preparation conditions and are therefore not precisely defined, it is consensus that these materials consist mainly of heptazine units connected by imino groups, with the formal composition $C_6N_9H_3$. Melon and related materials are currently extensively explored as photocatalysts for water splitting as well as for the oxidation of organic pollutants [21-27]. The heptazine molecule itself has been synthesized only relatively recently and its structure confirmed by X-ray diffraction [37]. The monomeric heptazine molecule reacts readily with water in the presence of light (see Section 5) and is therefore not routinely available.

The atomic structure, thermodynamic stability and electronic band structure of crystalline or polymeric materials consisting of heptazine building blocks were investigated with DFT calculations [38-42]. Cluster models of heptazine oligomers have been considered and the effects of corrugation on the electronic structure and the optical spectra have been explored [43,44]. The effects of chemical modifications of the monomers and doping of the polymers have been modelled with DFT calculations [45-48]. Structural aspects of hydrogen bonding of water with heptazine oligomers also were investigated [49-51].

The three chromophores pyridine, triazine and heptazine form hydrogen bonds with water as H-atom acceptors as illustrated in Fig. 1. The length of the hydrogen bond is 2.008 Å, 2.033 Å and 2.046 Å for pyridine, triazine and heptazine, respectively (calculated at the MP2 level) [52-55]. The calculations reveal that the nonbonding orbital of the N-atom involved in the hydrogen bonding with the water molecule becomes partially delocalized over the complex, indicating that the hydrogen bond exhibits some covalent bonding character [52-55].

### 3. Photooxidation of water

To provide a qualitative overview of the electronic absorption spectra of the three chromophore-water complexes, the vertical excitation energies of the seven lowest singlet excited states are collected in Table 1. For a more detailed discussion of the vertical excitation spectra, we refer to the original publications [52-55]. All excitations correspond to electronic transitions within the chromophores, since water does not absorb in this energy range. In the pyridine-water complex, the lowest $^1n\pi^*$ and $^1\pi\pi^*$ excited states are essentially degenerate with excitation energies of about 5.3 eV due to a blue shift of the $^1n\pi^*$ state of about 0.2 eV relative to isolated pyridine. In the triazine-water complex, the lowest $^1n\pi^*$ state is located at about 4.5 eV, about 1.0 eV lower than the lowest $^1\pi\pi^*$ excited state. In the heptazine-water complex, the lowest excited singlet state is a $^1\pi\pi^*$ state with an exceptionally low excitation energy of 2.6 eV, while the lowest $^1n\pi^*$ state is located at 3.7 eV. All these data refer to ADC(2) calculations with the cc-pVDZ basis set (see Refs. 52-54 for a description of the computational methods). The ADC(2) method tends to systematically overestimate excitation energies by about 0.2 – 0.3 eV; the actual excitation energies may therefore be lower by this amount.

*Table 1. Vertical excitation energies and oscillator strengths ( f ) of the seven lowest singlet excited states of the pyridine-$H_2O$, triazine-$H_2O$ and heptazine-$H_2O$ complexes computed with the ADC(2) method.*

|  | Py-$H_2O$ | | | Tr-$H_2O$ | | | Hept-$H_2O$ | | |
|---|---|---|---|---|---|---|---|---|---|
| state | character | energy (eV) | f | character | energy (eV) | f | character | energy (eV) | f |
| $S_1$ | $n\pi^*$ | 5.18 | 0.004 | $n\pi^*$ | 4.48 | 0.000 | $\pi\pi^*$ | 2.60 | 0.000 |
| $S_2$ | $\pi\pi^*$ | 5.23 | 0.040 | $n\pi^*$ | 4.60 | 0.005 | $n\pi^*$ | 3.72 | 0.000 |
| $S_3$ | $n\pi^*$ | 5.51 | 0.000 | $n\pi^*$ | 4.73 | 0.003 | $n\pi^*$ | 3.80 | 0.000 |
| $S_4$ | $\pi\pi^*$ | 6.59 | 0.020 | $n\pi^*$ | 4.77 | 0.000 | $n\pi^*$ | 3.89 | 0.000 |
| $S_5$ | $\pi\pi^*$ | 7.30 | 0.582 | $\pi\pi^*$ | 5.70 | 0.000 | $\pi\pi^*$ | 4.20 | 0.256 |
| $S_6$ | $\pi\pi^*$ | 7.41 | 0.645 | $\pi\pi^*$ | 7.08 | 0.002 | $\pi\pi^*$ | 4.22 | 0.252 |
| $S_7$ | $n\pi^*$ | 7.88 | 0.001 | $\pi\pi^*$ | 7.65 | 0.332 | $n\pi^*$ | 4.75 | 0.002 |

The computed oscillator strengths of the $^1n\pi^*$ states are very low for all complexes, see Table 1. These states can borrow intensity, however, from higher-lying allowed $^1\pi\pi^*$ excited states by vibronic coupling (Herzberg-Teller mechanism). The radiative transition to the lowest $^1\pi\pi^*$ excited state of heptazine (at 2.6 eV) is symmetry forbidden in the dipole approximation in the isolated chromophore ($D_{3h}$ symmetry). This transition may become weakly allowed by intramolecular vibronic coupling as well as by deformations of the heptazine building blocks from $D_{3h}$ symmetry in polymeric materials. In condensed materials, such as one-dimensional or two-dimensional organic polymers, low molecular oscillator strengths are sufficient for substantial light absorption. Guided by these considerations, we focus the attention in the following discussion on the lowest $^1\pi\pi^*$ excited state of the pyridine-$H_2O$ complex (absorption at 5.3 eV (234 nm)), the lowest $^1n\pi^*$ excited state of the triazine-$H_2O$ complex (absorption at 4.5 eV (276 nm)) and the lowest $^1\pi\pi^*$ excited state of the heptazine-$H_2O$ complex (absorption at 2.6 eV (477 nm)). While the smaller aromatic chromophores absorb in the UV (pyridine) or near UV (triazine), heptazine absorbs in the visible range of the spectrum.

In addition to the familiar excited states of the chromophores (which will be referred to as locally excited states in the following), there exist excited states of charge-transfer character in the hydrogen-bonded complexes, in which an electron is excited from either the $p_z$ orbital of water (nominally a π orbital in the planar systems) or one of the $p_x$, $p_y$ orbitals of water (nominally n orbitals) to one of the $\pi^*$ orbitals of the chromophore. These charge-transfer states are optically dark and are rather high in energy at the ground-state equilibrium geometries of the complexes. These states are therefore difficult to identify in the vertical electronic excitation spectrum. However, these charge-transfer states are dramatically stabilized in energy (by 2- 3 eV) when the proton of water involved in the hydrogen bond with the chromophore moves from water to the chromophore, neutralizing the electronic charge separation. Surprisingly, these charge-transfer states not only become the lowest excited states when their energy is optimized with respect to the nuclear coordinates, but even drop below the energy of the closed-shell electronic ground state (in other words, their excitation energy becomes negative). At their equilibrium geometries, these charge-transfer states represent neutral biradical states of the type pyridinyl-OH, triazinyl-OH or heptazinyl-OH, respectively (see Fig. 2). The $H_2O$ molecule has been oxidized in these biradical states by the abstraction of an H-atom.

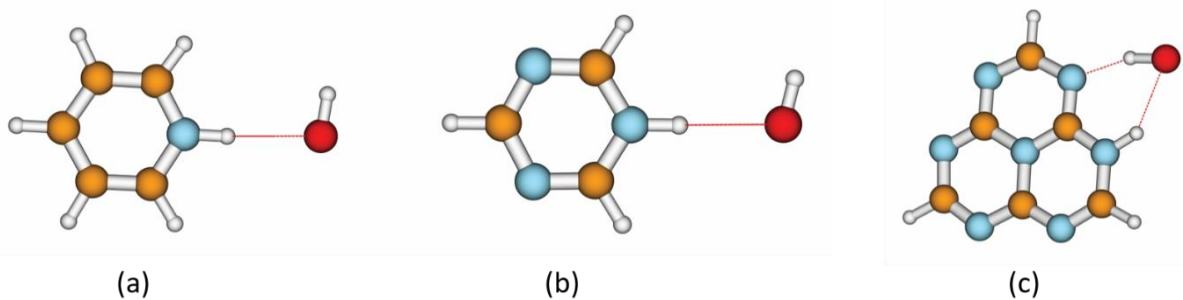

*Fig. 2. Geometries of the pyridinyl-OH (a), triazinyl-OH (b) and heptazinyl-OH biradicals. The geometry has been optimized in the $^1p_z\pi^*$ charge-transfer state at $R_{OH}$ = 2.2 Å.*

To illustrate the energetics of the water-oxidation reaction in these complexes, minimum-energy reaction paths were computed as so-called relaxed scans with the ADC(2) method. The driving coordinate for the relaxed scan was chosen as the bond length $R_{OH}$ of the OH bond of the water molecule which is involved in hydrogen bonding with the N-atom of the chromophore. For a fixed value of $R_{OH}$, the energy of a selected electronic state was optimized with respect to all other nuclear coordinates. Two such relaxed scans were computed, one by optimizing the energy of the electronic ground state of the complex, the other by optimizing the energy of the lowest charge-transfer state. Small values of $R_{OH}$ ($R_{OH} \approx$ 1.0 Å) represent the chromophore-water complex as shown in Fig. 1. Large values of $R_{OH}$ ($R_{OH} \approx$ 2.2 Å) represent a reduced chromophore radical (pyridinyl, triazinyl, heptazinyl) which is hydrogen bonded to a hydroxyl radical. The structures of these biradicals are shown in Figs. 2a-c.

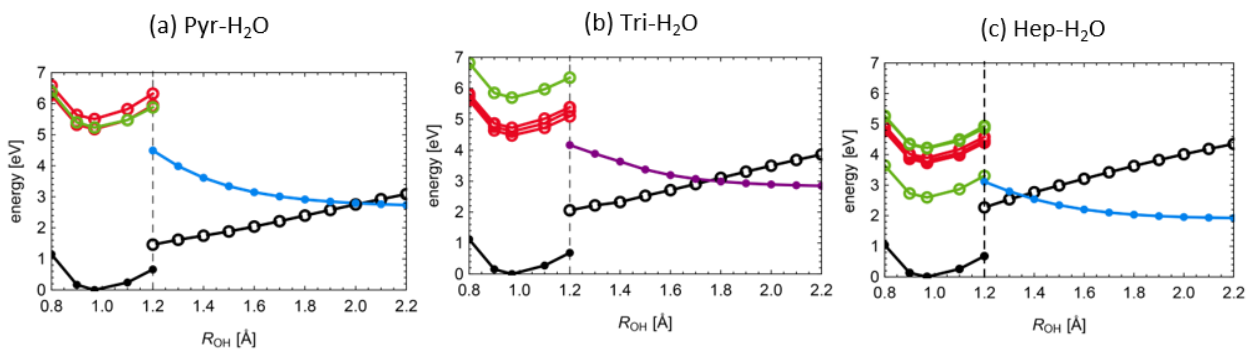

*Fig. 3. Energy profiles for the excited-state H-atom transfer reaction from water to the heterocycle for the pyridine-$H_2O$ (a), triazine-$H_2O$ (b) and heptazine-$H_2O$ (c) complexes, calculated with the ADC(2) method. Full dots indicate that the energy of this state was optimized. Open circles represent energies which were calculated at the optimized geometry of another electronic state. In the left part of the figures ($R_{OH}$ < 1.2 Å), the energy was optimized for the electronic ground state (black). In the right part of (a) and (c) ($R_{OH}$ > 1.2 Å), the energy of the $^1p_z\pi^*(CT)$ state (blue) was optimized, while in (b) the energy of the $p_{x,y}\pi^*(CT)$ state (violet) was optimized. The energy profiles of the locally excited $^1\pi\pi^*$ and $^1n\pi^*$ states are shown in green and red, respectively.*

The energy profiles of these relaxed scans are shown in Fig. 3. The $S_0$-optimized energy profiles are shown in the range 0.8 Å < $R_{OH}$ < 1.2 Å, while the energy profiles optimized for the charge-transfer states are shown in the range 1.2 Å < $R_{OH}$ < 2.2 Å. The vertical dashed line in Fig. 3 separates the two reaction paths. As explained above, we consider the lowest locally-excited $^1\pi\pi^*$ state and the corresponding charge-transfer state ($^1p_z\pi^*$) for the pyridine-$H_2O$ complex. For the triazine-$H_2O$ complex, we consider the lowest $^1n\pi^*$ state and the corresponding charge-transfer state ($^1p_{x,y}\pi^*$). For the heptazine-$H_2O$ complex, the lowest locally-excited $^1\pi\pi^*$ state and the corresponding $^1p_z\pi^*$ charge-transfer state are considered. The energy profiles of the corresponding triplet states were characterized for the pyridine-$H_2O$ and triazine-$H_2O$ complexes [53,54]. They are overall very similar to the singlet energy profiles and we refer to the publications [53,54] for details. It suffices to say that intersystem crossing from the excited singlet states to the triplet states generally is not a loss channel for the water oxidation reaction, since the H-atom transfer reaction can continue on the triplet surfaces after intersystem crossing [53,54].

In all three complexes, the potential-energy (PE) function of the lowest charge-transfer state is seen to cross the PE function of the $S_0$ state with increasing $R_{OH}$. This energy crossing is located at 1.70 Å in pyridine-$H_2O$, 1.75 Å in triazine-$H_2O$ and 1.35 Å in heptazine-$H_2O$. For the $^1p_{x,y}\pi^*$ charge-transfer state of A" symmetry in triazine-$H_2O$, the energy crossing with the $S_0$ state of A' symmetry is a symmetry-allowed crossing, giving rise to a symmetry-induced conical intersection [56] when out-of-plane vibrational modes are taken into account. For the $^1p_z\pi^*$ charge-transfer states of A' symmetry in pyridine-$H_2O$ and heptazine-$H_2O$, on the other hand, the energy crossing with the $S_0$ state is an avoided crossing in the one-dimensional picture of Fig. 3, which becomes an accidental same-symmetry conical intersection [56] when all nuclear degrees of freedom are taken into account. The nonadiabatic wave-packet dynamics at these conical intersections determines the yield of the water-oxidation reaction. If the wave packet on the charge-transfer PE surface follows the diabatic path (that is, stays on the charge-transfer PE surface at the crossing) a radical pair is formed. If, on the other hand, the wave packet follows the adiabatic path at the conical intersection (that is, switches to the $S_0$ surface at the crossing), the reaction is aborted and the complex relaxes to the minimum of the $S_0$ PE surface by intramolecular vibrational energy redistribution and energy transfer to the environment.

The connection of the PE surfaces of the locally excited states in the region of small $R_{OH}$ (chromophore-water complex) and large $R_{OH}$ (biradical) was explored by the computation of two-dimensional relaxed scans with the ADC(2) method, considering the distance between the H-atom donor (O) and the H-atom acceptor (N) as a second reaction coordinate. The energy of the lowest excited state of the considered symmetry was minimized with respect to the remaining nuclear coordinates, with the exception of the CH bond lengths. It was necessary to freeze these bond lengths to suppress undesired side reactions (such as photohydration, see below). The resulting two-dimensional PE surfaces are displayed in Figs. 4a-c.

The PE surfaces of all three systems exhibit two minima which are separated by a saddle point (marked by a circle). The left minimum represents the equilibrium geometry of the locally excited state in the Franck-Condon region of the chromophore-water complex. The right minimum represents the equilibrium geometry of the corresponding biradical. At the saddle point, the character of the electronic state changes from locally excited (left) to charge transfer (right). The saddle points reflect the existence of nearby conical intersections (not visible in Fig. 4). Because the locally excited state and the charge-transfer states have the same symmetry, these are accidental same-symmetry conical intersections.

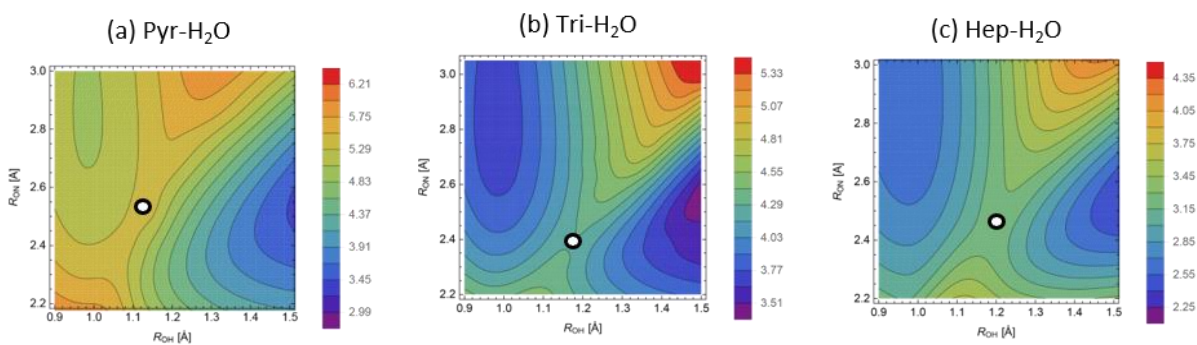

*Fig. 4. Relaxed potential-energy surfaces of the lowest excited state of the pyridine-$H_2O$ (a), triazine-$H_2O$ (b) and heptazine-$H_2O$ (c) complexes, calculated with the ADC(2) method. For fixed $R_{OH}$ and $R_{ON}$, the energy of the excited state has been optimized with respect to all other nuclear coordinates. The energy minimum for small $R_{OH}$ represents the Franck-Condon minimum of the lowest locally excited singlet state. The energy minimum for large $R_{OH}$ represents the biradical of the corresponding symmetry (A' in (a) and (c), A'' in (b)). The circle indicates the saddle point. The color code gives the energies (in eV) with respect to the energy minimum of the electronic ground state.*

On the $^1$A' PE surface of the pyridine-$H_2O$ complex, the estimated energy of the saddle point for H-atom transfer is 5.03 eV (relative to the $S_0$ energy minimum), which is about 0.4 eV higher than the energy minimum of the locally excited $^1\pi\pi^*$ state, but 0.4 eV lower than the vertical excitation energy of the $^1\pi\pi^*$ state. The H-atom transfer reaction is thus barrierless with respect to the vertical excitation energy of the complex. Even with respect to the minimum of the $^1\pi\pi^*$ surface, the barrier for H-atom transfer is only a few tenth of an electron volt higher than the zero-point energy, which implies rapid H-atom tunneling. Rather efficient H-atom transfer is therefore expected in photoexcited pyridine-$H_2O$. This expectation is confirmed by a recent experiment for pyridine-$(H_2O)_n$ clusters in a molecular beam, in which the photoinduced H-atom transfer reaction was confirmed by the detection of the pyridinyl radical with mass spectrometry [29].

On the $^1$A'' PE surface of the triazine-$H_2O$ complex, the saddle point is found at 4.31 eV, 0.6 eV above the energy minimum of the locally excited $^1n\pi^*$ state, but 0.17 eV below the vertical excitation energy of this state. The H-atom transfer reaction is thus barrierless with respect to the vertical excitation energy of the $^1n\pi^*$ state. The somewhat higher barrier with respect to the $^1n\pi^*$ energy minimum (compared with the $^1\pi\pi^*$ state of pyridine) reflects the larger vibrational stabilization energy of the locally excited $^1n\pi^*$ state of triazine.

On the $^1$A' PE surface of heptazine-$H_2O$, the energy of the saddle point is 3.30 eV, about 0.8 eV above the energy minimum of the locally excited $^1\pi\pi^*$ state and 0.7 eV above the vertical excitation energy of this state. Notable features of the heptazine-$H_2O$ complex are the low vertical excitation energy of the lowest locally excited $^1\pi\pi^*$ state (2.60 eV) and the rather low vibrational stabilization energy of this state. In comparison with the pyridine-$H_2O$ and triazine-$H_2O$ complexes, the saddle point for H-atom transfer is higher than the vertical excitation energy in heptazine-$H_2O$ and the H-atom transfer will require barrier tunneling. These findings are in qualitative accordance with the experimental observation that hydrogen evolution upon irradiation of melon at 500 – 600 nm is significantly less efficient than upon irradiation at 420 nm [20,57]. In the latter case, $^1n\pi^*$ locally excited states of heptazine are populated. After internal conversion to the lower-lying $^1\pi\pi^*$ state, substantial vibrational excess energy is available for barrier crossing on the $^1\pi\pi^*$ PE surface, resulting in efficient H-atom transfer.

## 4. Photodissociation of heterocyclic radicals

The pyridinyl, triazinyl and heptazinyl radicals are planar systems. The NH bond length is about 1.01 Å in all three radicals. The singly occupied orbital is a π orbital. In all three systems, a so-called σ* orbital which is localized on the NH group is among the lowest unoccupied orbitals. Like the σ* orbitals of pyrrole or indole [58], this orbital is a diffuse Rydberg orbital which is antibonding with respect to the NH bond. Moreover, the density of the σ* orbital is mostly localized outside the aromatic ring. Due to these features, the occupation of the σ* orbital provides the driving force for the dissociation of the NH bond [58].

The pyridinyl and triazinyl radicals possess a single $^2\pi\pi^*$ excited state below 2.0 eV, while the heptazinyl radical possesses two such low-lying $^2\pi\pi^*$ states. In pyridinyl and triazinyl, the reactive $^2\pi\sigma^*$ state is the second lowest state, while it is the third lowest state in heptazinyl. The vertical excitation energies of the $^2\pi\sigma^*$ state are 2.05 eV, 2.85 eV and 3.21 eV for pyridinyl, triazinyl and heptazinyl, respectively, at the UADC(2)/aug-cc-pVDZ level. All three radicals exhibit a dense spectrum of $^2\pi\pi^*$ and $^2n\pi^*$ excited states above 3.0 eV. The low-lying $^2\pi\pi^*$, $^2n\pi^*$ and $^2\pi\sigma^*$ excited states have very small oscillator strengths, but can borrow intensity from higher-lying bright states by vibronic coupling. The bright $^2\pi\pi^*$ excited states of the radicals are slightly higher in energy than the corresponding bright $^1\pi\pi^*$ states of pyridine, triazine and heptazine [52,54,55].

Experimental information on the excitation energies exists only for the pyridinyl radical, which was prepared by Jouvet and coworkers in a molecular beam [29]. The vibronically resolved resonant multi-photon ionization (REMPI) spectra of four excited states of the pyridinyl radical were detected in the range 430 - 300 nm and assigned by comparison with CC2 electronic-structure calculations [29].

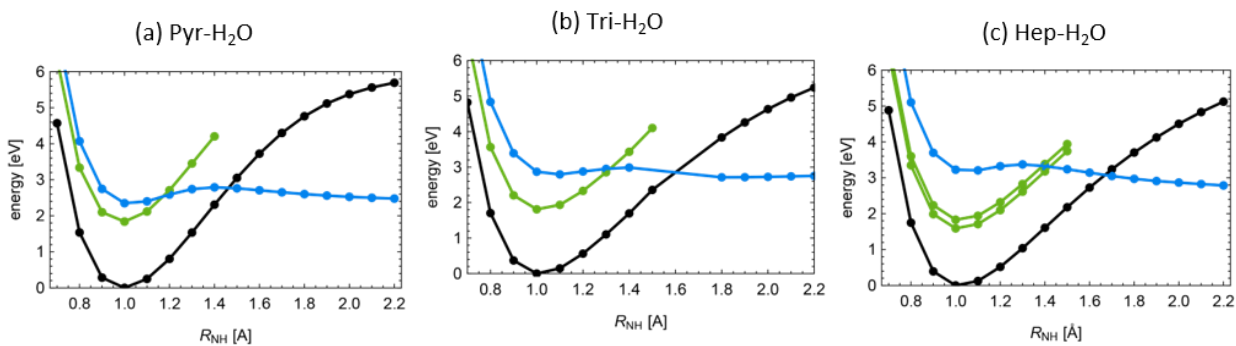

Fig. 5. Potential-energy functions of the pyridinyl radical (a), the triazinyl radical (b) and the heptazinyl radical (c) along the NH stretching coordinate, calculated with the ADC(2) method. Black: electronic ground state; blue: $^2\pi\sigma^*$ state; green: $^2\pi\pi^*$ state. For clarity, $^2n\pi^*$ states and $^2\pi\pi^*$ states with vertical excitation energies above the $^2\pi\sigma^*$ state are not shown.

The energy profiles of the ground state and the lowest excited states of the three radicals were computed as rigid scans (that is, varying the NH stretching coordinate, while keeping all other internal coordinates fixed). The PE functions obtained for the electronic ground state ($D_0$), the lowest $^2\pi\pi^*$ state(s) and the lowest $^2\pi\sigma^*$ state are displayed in Figs. 5a-c. While the PE function of the $D_0$ state and those of all $^2\pi\pi^*$ and $^2n\pi^*$ states are bound with respect to the NH stretching coordinate, the PE function of the $^2\pi\sigma^*$ state is dissociative with a very low dissociation energy. For pyridinyl and triazinyl, the dissociation energy was benchmarked with the CASPT2 method and a value of 2.0 eV was obtained for both radicals [52,53]. For heptazinyl, the dissociation threshold was computed with the CCSD coupled-cluster method and the same value (2.0 eV) was obtained [55]. The dissociative PE function of the $^2\pi\sigma^*$ state crosses the PE function of the $D_0$ state at 1.50 Å in pyridinyl, 1.60 Å in triazinyl and 1.70 Å in heptazinyl, respectively. These energy crossings are symmetry-allowed conical intersections (the $D_0$ state is of $^2A''$ symmetry, while the $^2\pi\sigma^*$ state is of $^2A'$ symmetry in the $C_s$ symmetry group). In addition, there is a crossing of the energy of the $^2\pi\sigma^*$ state with the energy of the lowest $^2\pi\pi^*$ state(s), see Fig. 5.

The $^2\pi\sigma^*$ state of the radicals can be either excited directly with visible light with the help of intensity borrowing from higher electronic states, or the $^2\pi\sigma^*$ state can be populated via fast radiationless transitions after excitation of higher bright electronic states. The photodissociation dynamics of the pyridinyl radical was recently studied by wave-packet dynamics simulations for reduced-dimensional models [59]. It was shown that efficient photodissociation through several conical intersections is possible for these reduced-dimensional models [59]. The photodetachment of the pyridinyl radical was also experimentally detected by irradiation of pyridinyl-water clusters in a molecular beam [29].

The photodissociation of the radicals regenerates the original chromophores and thus closes the catalytic cycle. Overall, the water molecule which is hydrogen-bonded to the chromophore is decomposed into hydrogen and hydroxyl radicals in a biphotonic reaction. Two photons are sequentially absorbed by a single chromophore and each photon promotes a hydrogen-transfer or hydrogen-detachment reaction. The catalytic cycle can alternatively be closed by a dark radical recombination reaction [52-55]. Two radicals, which were generated by a photoinduced hydrogen-abstraction reaction in two different chromophore-water complexes, can recombine to form $H_2$ in an exothermic reaction, thereby regenerating two chromophore molecules. In the first scenario, four photons are needed to generate two free hydrogen atoms, whereas in the second scenario, two photons generate one $H_2$ molecule. The enthalpy of the two free hydrogen atoms generated in the first scenario is about 4.4 eV (100 kcal/mol) higher than the enthalpy of the $H_2$ molecule generated in the second scenario.

## 5. Reactivity of OH radicals with the reduced chromophores

The results discussed so far suggest that water molecules can be decomposed into hydrogen atoms and hydroxyl radicals by a rather simple biphotonic process, using N-heterocycles as chromophores and photocatalysts. The electron-driven proton-transfer reactions in excited states are characterized by low energy barriers of the order of a few tenths of an electron volt, in sharp contrast to the energy barriers of at least 2.0 eV encountered in the water oxidation reaction in the electronic ground state [35,51]. Rather than being barrier-controlled, the efficiencies of the excited-state H-atom transfer and H-atom detachment reactions are controlled by the ultrafast nonadiabatic dynamics at conical intersections. While so far no quantitative theoretical studies of the dynamics at these conical intersections have been performed, there exists experimental evidence that OH radicals are readily generated when polymeric carbon nitride materials, in particular melon, are irradiated with polychromatic

UV/vis light [59] or with monochromatic laser light [61]. The OH radicals have been detected directly by ESR spectroscopy [62] or indirectly by their reaction with OH radical scavengers and the detection of the fluorescence of the reaction products [60,61]. Nascent hydrogen radicals generated by the irradiation of dilute suspensions of melon in water also have been detected, using the triiodide anion as a spectroscopic marker [61]. The joint detection of H and OH radicals [61] confirms that water molecules can indeed be decomposed homolytically into hydrogen and hydroxyl radicals by UV/vis radiation with N-heterocyclic photocatalysts. In fact, the challenge of photocatalytic water-splitting may be less the efficiency of the generation of radicals from water, but the control of the reactivity of the highly reactive OH radicals. Importantly, reactions of the OH radicals with the precious photocatalyst must be minimized.

The photohydration reaction of pyridine in aqueous solution is an instructive example. It is well known that UV irradiation of aqueous solutions of pyridine yields a product which exhibits an absorption maximum at 360 nm and slowly reverts back to pyridine in the dark [63-65]. This photoproduct was identified as the hydrate of pyridine. As is well known, the $^1\pi\pi^*$ excited states of the six-membered heterocycles pyridine, pyrazine, pyrimidine and triazine are prone to ultrafast internal conversion to the electronic ground state via $^1\pi\pi^*$-$S_0$ conical intersections which are accessible by out-of-plane puckering of the aromatic ring [66-68]. This ultrafast photochemistry yields the hot ground state of these molecules or hot valence isomers. These highly excited species are able to react with water molecules of the aqueous environment, yielding the respective photohydrates [68]. This photohydration reaction is also well established for the pyrimidine bases uracil and thymine and is known to be a significant source of lesions in DNA [69,70]. For all these six-membered heterocycles, the photohydrates are metastable and dehydrate to the original chromophore in dark reactions on long time scales [63-65,69,70].

Interestingly, the photohydrated chromophore may alternatively be formed by a follow-up reaction of the photoinduced homolytic water-splitting reaction in chromophore-water complexes [71]. In the simplest case of the pyridinyl-OH biradical, the OH radical may attach to one of the CH groups of pyridinyl and form a covalent bond, which corresponds to a nonadiabatic electronic transition from the biradicalic structure to a closed-shell electronic structure. The molecular structure of this pyridinyl hydroxide is shown in the second panel of Fig. 6a. This structure can rearrange by the transfer of a hydrogen atom from the hydroxyl group to the NH group via a low energy barrier, which results in the closed amino-aldehyde from of the photohydrate shown in the third panel of Fig. 6a. This cyclic structure can finally rearrange to the more stable open amino-carbonyl structure shown in the fourth panel of Fig. 6a. The energies of these photohydrates relative to the energy minimum of the pyridine-$H_2O$ complex computed at the MP2/cc-pVDZ level are of the order of 23 kcal/mol, see Fig. 6a. The energy of the open amino-aldehyde (fourth panel) is somewhat higher in the gas phase than the energy of the pyridinyl hydroxide (second panel). This energetic order is likely to change in aqueous solution due to the larger solvation energy of the open form. It is seen that all pyridine hydrates are thermodynamically metastable and therefore will eventually relax back to pyridine and water, as is observed experimentally [63-65]. The photohydration of pyridine is thus a loss channel for the water-splitting reaction, but does not represent a hazard for the photostability of the photocatalyst.

The corresponding structures for photohydrated triazine are shown in Fig. 6b. The energies of the triazine photohydrates are of the order of 12 kcal/mol higher than the energy of the triazine-water complex. These structures also are metastable and the triazine photohydrates will eventually decay back to triazine and water. The calculated structures of photohydrated heptazine are displayed in Fig. 6c. In this case, the open amino-aldehyde form is the most stable structure and, remarkably, is lower in energy than the heptazine-water complex. The open form of the photohydrate of heptazine is thus thermochemically stable. This result is in agreement with the experimental observation that heptazine hydrolyzes spontaneously and irreversibly whenever traces of water are present [37].

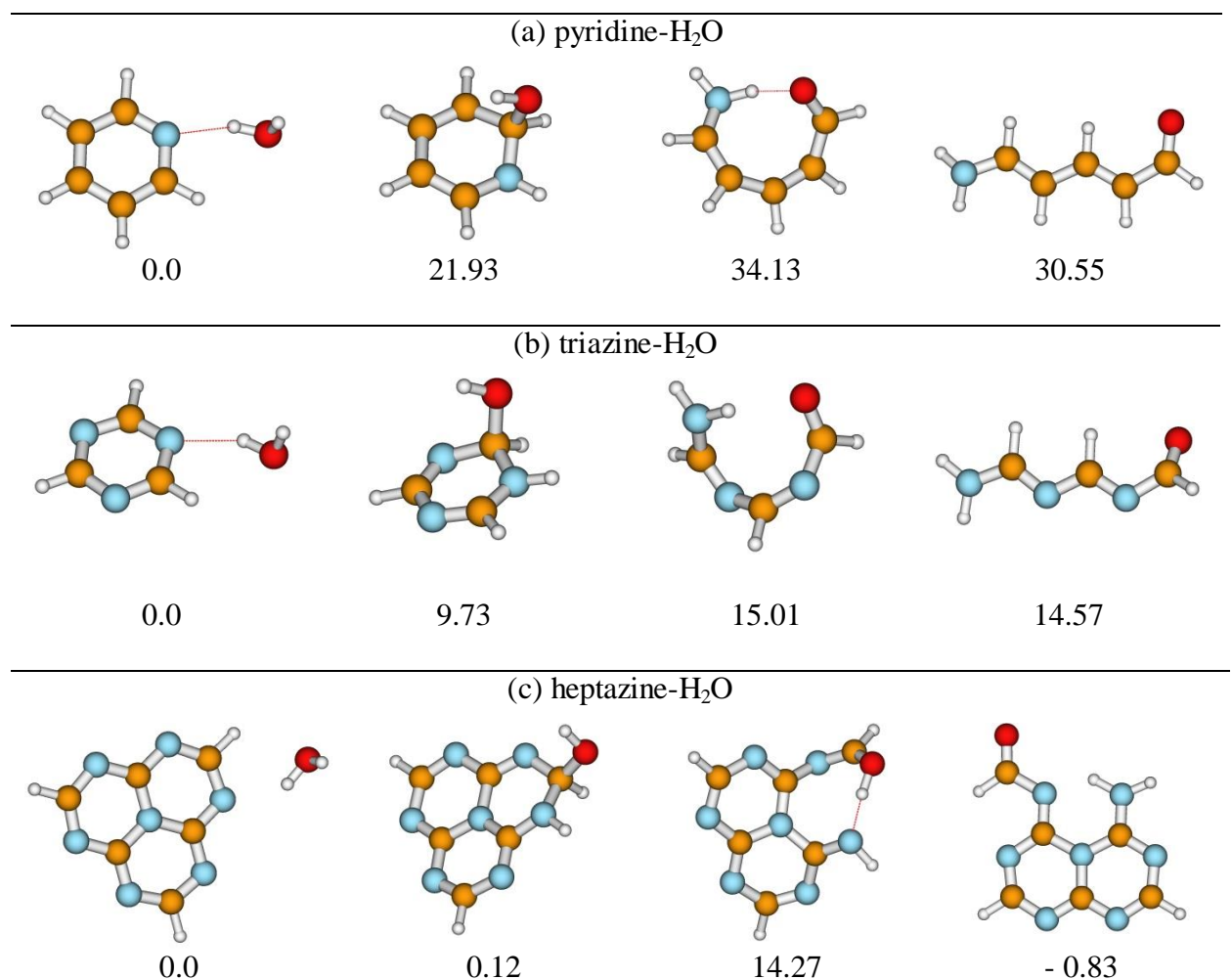

*Fig. 6. Structures and relative energies (in kcal/mol) of photohydrates of pyridine (a), triazine (b) and heptazine (c), computed at the MP2/cc-pVTZ level.*

There is an essential difference in the mechanism of the photohydration reactions of the single-ring heterocycles (pyridine and triazine) and the fused-ring heterocycle heptazine. As discussed above, the dominant photohydration mechanism in pyridine-$H_2O$ is ultrafast internal conversion through ring-puckering conical intersections, followed by the reaction of the hot ground state with a water molecule [63-65]. In the fused heterocycle heptazine, on the other hand, ring-puckering in excited states is suppressed and the excited states of heptazine (as well as the excited states of the heptazine-based polymer melon) have long lifetimes with high fluorescence quantum yields [21-27]. This implies that photohydration of heptazine can only occur as a follow-up reaction of photoinduced homolytic water splitting, that is, by the reaction of the photogenerated hydroxyl radical with the heptazinyl radical via the intermediate structures shown in Fig. 6c. The spontaneous hydrolysis of heptazine (in the presence of light) is thus a witness of heptazine's ability to split water with visible light.

The photohydration reactions represented by the intermediate structures in Fig. 6 occur via the attack of an OH radical on the aromatic ring at a CH group in α position to the reduced N-atom. The formation of the carbonyl group provides the energy needed for the dissociation of the OH radical. In the polymer melon, however, these CH groups are replaced by CN bonds, which are robust with respect to hydroxylation. This explains the amazing photostability and robustness of melon against photohydrolysis [21-24] in sharp contrast to monomeric heptazine [37].

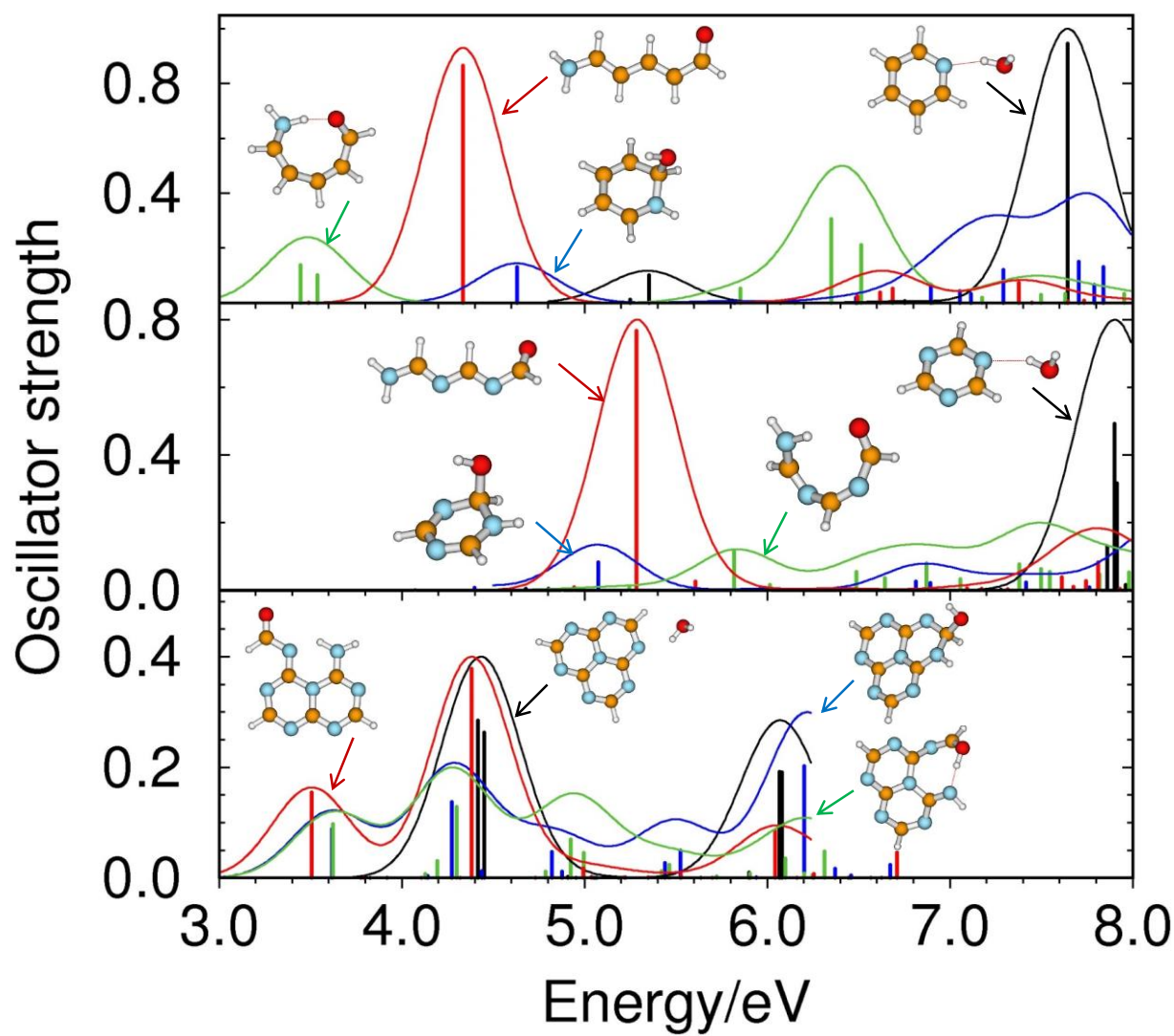

*Fig. 7. Absorption spectra of the chromophore-water complex (black), the hydroxylated hydrogenated radical (blue), the closed form of the amino-aldehyde (green) and the open form of the amino-aldehyde (red). The chromophores are pyridine (upper panel), triazine (middle panel) and heptazine (lower panel). The corresponding structures are shown as insets. The computed stick spectra were convoluted with a Gaussian function of 0.5 eV FWHM.*

The various photohydrates shown in Fig. 6 can be experimentally identified via their absorption or luminescence spectra. In Fig. 7 we present an overview of the calculated electronic absorption spectra (vertical excitation energies and oscillator strengths calculated with the ADC(2) method) of the species shown in Fig. 6. The computed electronic stick spectra are convoluted with a Gaussian of 0.5 eV FWHM to simulate typical absorption profiles in solution. The spectra of different species are represented by different colors which are assigned by the insets. It should be noted that vibronic intensity borrowing effects are not included in these calculations; vibronically induced weak transitions are therefore missing in these spectra.

Beginning with the pyridine-$H_2O$ complex (upper panel in Fig. 7), the absorption band of the lowest $^1\pi\pi^*$ state of pyridine-$H_2O$ (black) is predicted at about 5.4 eV with a moderate oscillator strength, see Fig. 7. The lowest absorption band of pyridinyl hydroxide is found at 4.6 eV (blue), slightly below the lowest absorption band of pyridine-$H_2O$. The closed form of the amino-aldehyde exhibits the lowest absorption band (near 3.5 eV, red). The open form of the amino-aldehyde is predicted to absorb much more strongly than the other species near 4.3 eV (288 nm, green). The blue shift with respect to the experimentally observed strong absorption band at 360 nm [64,65] is due to the lack of solvation in the present calculations.

The corresponding absorption spectra of the triazine-$H_2O$ complex and its photohydrates are displayed in the middle panel of Fig. 7. While several $^1\pi\pi^*$ states of triazine absorb strongly near 7.9 eV (black), the absorption lines of the $^1n\pi^*$ states near 4.8 eV are not visible due to their low oscillator strengths. The triazinyl hydroxide exhibits a moderate absorption near 5.0 eV (248 nm, blue), whereas the closed form of the amino-aldehyde absorbs broadly above 5.8 eV (red). As found for the pyridine hydrates, the open form of the amino-aldehyde absorbs much more strongly than the other photohydrated species. This strong absorption band is predicted near 5.2 eV (240 nm) in the gas phase, at a somewhat shorter wavelength than the corresponding pyridine hydrate.

The spectra of the heptazine photohydrates are shown in the lower panel of Fig. 7. In this case, the heptazine-$H_2O$ complex (black), the heptazinyl hydroxyde (blue) and the amino-aldehyde (red) all absorb near 4.5 eV (275 nm). Heptazinyl hydroxide and the amino-aldehyde exhibit additional weaker absorption bands near 3.5 eV (355 nm). The thermodynamically stable amino-aldehyde does not absorb at wavelengths longer than 400 nm. It is therefore colorless and cannot be converted back to heptazine and water by the absorption of visible light. This explains the irreversible hydrolysis of heptazine in the presence of water and visible light [37].

## 6. Conclusions

Throughout the current literature, the mechanism of (scacrificial) hydrogen evolution with polymeric carbon nitrides and UV/vis light is interpreted in terms of exciton formation in the polymers, followed by exciton dissociation, separation of mobile charges and finally the reaction of electrons and holes with water molecules at solid-liquid interfaces [20-27]. In view of large exciton dissociation energies [72-74], significant polaron stabilization energies [75,76] and comparatively low charge carrier mobility [74] in organic polymers, it appears questionable whether this paradigm, which was developed for the interpretation of photoelectrochemical water splitting with transition-metal oxides [3,4], is applicable for carbon nitride materials such as triazine-based covalent organic frameworks or heptazine-based graphitic carbon nitrides. We have sketched in this perspective an alternative paradigm which emerged from the exploration of the specific photochemistry of aromatic heterocycles in hydrogen-bonded complexes with water molecules using first-principles electronic-structure computational methods. The computational results provide evidence that the chromophores pyridine, triazine and heptazine, when photoexcited to low-lying $^1\pi\pi^*$ or $^1n\pi^*$ electronic states, can abstract a hydrogen atom from hydrogen-bonded water molecules by an electron-transfer reaction (from the water molecule to the chromophore, where the electron fills the hole in one of the π orbitals generated by the photoexitation) which is followed by a proton-transfer reaction from the water molecule to the chromophore. The calculations predict that the energy barriers involved in the excited-state PCET reactions are generally of the order of a few tenth of an electron volt, in sharp contrast to H-atom transfer barriers in the electronic ground state, which typically are at least 2.0 eV. The excited-state proton transfer reaction is therefore expected to be fast and the preceding electronic charge separation should be neutralized on a time scale of femtoseconds. Considering the low mass of the proton compared to the effective masses of most other vibrational degrees of freedom, it is likely that the proton-transfer reaction will be more efficient than competing loss reactions, such as, for example, radiative or radiationless decay to the electronic ground state.

In the second step of the biphotonic water-splitting reaction, the excess hydrogen atom is photodetached from the aromatic radical via a direct and fast (nonstatistical) photodissociation reaction. The calculations predict the existence of a low-lying dissociative $^2\pi\sigma^*$ state in all three N-hydrogenated radicals considered herein. This $^2\pi\sigma^*$ state drives a photodetachment reaction which is very similar to the well-established $^1\pi\sigma^*$ photochemistry of closed-shell aromatic molecules with acidic groups, such as pyrrole, indole or aniline [77-81]. The photodetachment of the excess hydrogen atom from the aromatic radial regenerates the chromophore and thus closes the catalytic cycle. Overall, the hydrogen-bonded water molecule is decomposed into H and OH radicals by two sequential photochemical reactions.

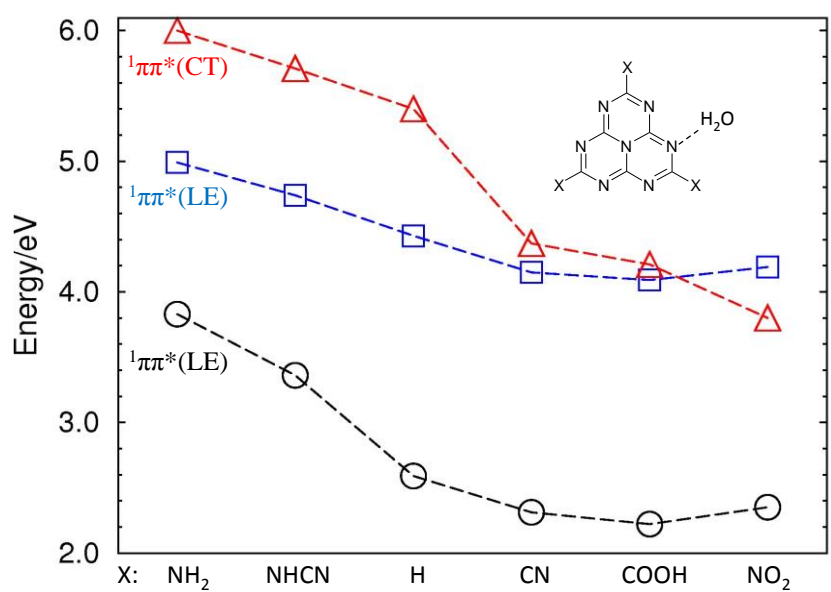

Fig. 8. *Vertical excitation energies of the lowest locally excited $^1\pi\pi^*$ states and the $^1p_z\pi^*$ charge-transfer state of complexes of substituted heptazine chromophores with a water molecule. X = H stands for heptazine.*

The observed systematic increase of the efficiency of hydrogen evolution with the nitrogen content of the heterocycles pyridine, pyrimidine and triazine in covalent organic frameworks [17,18] provides strong evidence for a crucial role of heterocyclic N-atoms in the photoinduced water-splitting reaction. Can this paradigm be further validated by spectroscopic or kinetic measurements? A straightforward option is chemical substitution of the triazine or heptazine chromophores at the three CH groups. As an example, we have computed the energies of the lowest two locally excited $^1\pi\pi^*$ states and the $^1p_z\pi^*$ state of heptazine derivatives which are triply substituted by $NH_2$, NHCN, CN, COOH and $NO_2$ groups, respectively. The results are displayed in Fig. 8. It is seen that the electron donating groups ($NH_2$, NHCN) increase the energies of both the locally exited states and the charge-transfer state. The electron withdrawing groups (CN, COOH, $NO_2$), on the other hand, have a minor effect on the vertical excitation energies of the locally exited states, but stabilize the energy of the charge-transfer state by 1.0 – 1.3 eV. This significant stabilization of the charge-transfer state relative to the locally excited states will result in a substantial lowering of the barrier for the H-atom transfer reaction. This implies that the photoinduced H-atom abstraction reaction from water will compete more efficiently with excited-state quenching reactions in the derivatives with electron withdrawing groups, which should result in enhanced efficiency of $H_2$ and $H_2O_2$ evolutions. In addition, asymmetric substitution of heptazine should be considered, which relaxes the symmetry selection rules which render the HOMO to LUMO transition dipole forbidden. Asymmetrically substituted heptazine derivatives should therefore be superior chromophores in the visible range of the spectrum.

Table 2. Vertical energies of the lowest excited singlet states of complexes of heptazine with selected solvent molecules, computed with the ADC(2) method.

| solvent | $1^1\pi\pi^*$ | $1^1n\pi^*$ | $2^1\pi\pi^*$ | $^1p_z\pi^*$(CT) |
|---|---|---|---|---|
| $H_2O$ | 2.59 | 3.76 | 4.43 | 5.40 |
| MeOH | 2.59 | 3.75 | 4.41 | 4.65 |
| 4-MBA[a] | 2.59 | 3.75 | 4.40 | 4.08 |
| MNAH[b] | 2.60 | 3.77 | 4.42 | 3.69 |
| TEOA[c] | 2.59 | 3.77 | 4.40 | 3.53 |

[a] (4-methylphenyl)methanol
[b] 1-methyl-1,4-dihydronicotinamide
[c] triethanolamine

Another prediction of the proposed paradigm is the dependence of the photoreactivity on the oxidation potential of the solvent molecules which serve as electron donors. The vertical excitation energies of the lowest three locally excited states and the lowest charge-transfer state are given in Table 2 for the hydrogen-bonded complexes of heptazine with a single water, methanol, 4-methylbenzoic acid (4-MBA), 1-methyl-1,4-dihydronicotinamide (MNAH) and triethanylamine (TEOA) molecule. It is seen that the vertical excitation energies of the locally excited states are not affected by the hydrogen bonding with the solvent molecule, while the energy of the $^1p_z\pi^*$ charge-transfer state decreases by nearly 2.0 eV from the complex of heptazine with $H_2O$ (5.40 eV) to the complex of heptazine with TEA (3.53 eV). In the complexes of heptazine with 4-MBA, MNAH and TEA the energy of the charge-transfer state is seen to be below the energy of the dipole-allowed $^1\pi\pi^*$ state, which indicates a barrierless excited-state H-atom transfer reaction from this state. In the heptazine-water complex, on the other hand, the energy of the charge-transfer state is nearly 1 eV higher than the energy of the dipole-allowed $^1\pi\pi^*$ state. Table 2 shows that the photooxidation of amines and alcohols is so much easier than the photooxidation of water that the demonstration of hydrogen evolution with the former solvents is of little relevance for the water oxidation reaction.

There is nevertheless computational as well as experimental evidence that the light-driven homolytic dissociation of water with N-heterocycles can be quite efficient. The commonly observed formation of photohydrates by N-heterocycles is a signature of water splitting, as discussed above. The challenge of developing efficient hydrogen generation with carbon nitrides may be less the decomposition of water as such, but rather the controlled recombination of the photogenerated radicals to the desired closed-shell products ($H_2$ and $H_2O_2$). This conclusion implies that investigations of hydrogen evolution with sacrificial reagents, such as amines and alcohols, provide little insight into the essentials of the problem. The sacrificial reagents either serve as electron donors (which implies that water is not oxidized at all) or alternatively as OH radical scavengers (which implies that the true challenge, the controlled recombination of OH radicals, is obscured).

In the opinion of the authors, priority should be given at the current stage of development of solar water splitting to the clarification of the fundamental mechanisms of the reaction. Water splitting should be perceived as a physical chemistry problem rather than a materials science problem. For example, irradiation of the photocatalyst with a tunable monochromatic laser of tunable intensity can provide much more insight into the molecular mechanisms of the reaction than irradiation with a standard polychromatic lamp or a sunlight simulator. Once the microscopic mechanisms of the decisive reactions are fully understood, the materials and the co-catalysts can be optimized in a more efficient manner than currently is the case.

## Acknowledgments

The research summarized in this article was supported by the DFG cluster of Excellence "Munich Centre for Advanced Photonics" (MAP). J. E. was partially supported by the International Max-Planck Research School of Advanced Photon Science (IMPRS-APS). A. L. S. acknowledges support by the Alexander von Humboldt Research Award.